\documentclass[a4paper,11pt]{article}
\usepackage{jcappub} 
\usepackage{lineno}
\usepackage{amssymb}
\usepackage{multirow}
\usepackage[dvipsnames]{xcolor}
\usepackage[normalem]{ulem}

\newcommand{\aap}{Astron.\ Astrophys.}

\newcommand{\prd}{Phys. Rev. D}
\newcommand{\prl}{Phys. Rev. Lett.}
\newcommand{\apj}{Astrophys.\ J.}

\newcommand{\jcap}{J. Cosmology \& Astropaticles}
\newcommand{\pasp}{Publ. Astron. Soc. Pacific}

\newcommand{\nat}{Nature}
\newcommand{\ssr}{Sp. Sci. Rev.}

\newcommand{\Eqg}{E_\text{QG}}
\newcommand{\Epl}{E_\text{Pl}}
\newcommand{\lf}{\text{lf}}
\newcommand{\cin}{\text{ci1}}
\newcommand{\cia}{\text{ci2}}
\newcommand{\jp}{JP08}
\newcommand{\ac}{A24}
\newcommand{\diff}{\mathrm{d}}

\title{\boldmath Constraining the curvature-induced quantum gravity scales via gamma-ray bursts}

\author[1]{D. D. Ofengeim\note{Corresponding author.}}
\author{and T. Piran}
\affiliation{Racah Institute of Physics, The Hebrew University, Jerusalem 91904, Israel}

\emailAdd{ddofengeim@gmail.com}

\abstract{
We constrain the parameters that govern curvature-induced quantum gravity time-of-flight (TOF) effects. These TOF delays, which occur due to modified dispersion relations of particles in 
the vacuum, could be a phenomenological signature of quantum gravity. Gamma-ray bursts (GRBs), short, high-energy events from distant galaxies, offer a unique opportunity to impose observational limits on TOF delays and, by extension, on the energy scales of quantum gravity. Using the standard Jacob-Piran relation, which assumes a locally-flat spacetime, the analysis of quantum gravity-induced TOF effects establishes a lower limit of approximately $10 \Epl$ on the energy scale of these effects. However, curvature-induced quantum gravity effects may introduce additional contributions. From current GRB observations, we find that, at a 95\% credibility level, in the symmetry-deformed scenario, curvature-induced TOF effects may only arise at energies above $0.04 \Epl$. If we consider only curvature-induced effects, this limit is an order of magnitude stronger. Observing more GRBs at different redshifts could improve the constraints on the curvature-induced QG phenomena. However, given the capabilities of current telescopes and the current understanding of GRBs, it is unlikely that these constraints will be significantly extended beyond the present level.
}

\begin{document}
\maketitle
\flushbottom

\section{Introduction}
\label{sec:intro}

One of the potential phenomenological manifestations of quantum gravity (QG) is a modification of the in-vacuum dispersion relation, i.e., breaking or deformation of the Lorentz invariance of the speed of massless particles \cite{Addazi+PrPNP2022}. Microscopically, it corresponds to the treatment of the quantum vacuum of the gravitational field as a dispersing medium, which can be polarized by a by-passing particle, with the corresponding back-reaction. The modified dispersion relation (MDR) leads to the time-of-flight (TOF) delay of photons with different energies. It is a cornerstone of astrophysical tests for QG phenomenology \cite{Amelino+Nat1998}: roughly speaking, detecting two photons of different energies from the same event, one can consider the duration of this event as an upper limit for the QG TOF delay between these photons. This allows us to set a lower limit on the energy scale $\Eqg$ at which the QG effects on the dispersion relation become significant.

Gamma-ray bursts (GRBs) are the best transients for such purposes. 
{They}
are short-duration high-energy transients, that are variable on short time scales.  They are cosmologically distant, so the QG TOF delay would accumulate up to relatively large values. 
Thus, QG TOF delay, if exists, would be relatively significant in GRB observations.
While there is no explicit detection of QG effects yet, current studies of TOF delays in individual GRBs allow to set limits on the quantum-gravity energy scale $\Eqg$ up to ten times the Planck energy $\Epl = 1.22\times 10^{19}\,$GeV \cite{Vasileou+PRD2013,MAGIC_190114LIV,PiranOfengeimPRD2024,LHAASO_221009LIV}. The application of the TOF method to GRBs suffers from systematic uncertainties, mainly due to unknown intrinsic lags between photons of different energies, which are difficult to take into account (e.g., \cite{Ellis+ApP2006,Ellis+PRD2019,Vardanyan+PRD2023}). However, we can overcome those by combining observations from different GRBs at different cosmological distances and from different emission regions with GRBs.

The standard formula for the QG TOF delay $\Delta t(\Delta E, z)$ as a function of the energy lag
\footnote{We consider only linear QG effects, with $\Delta t \propto \Delta E$. The quadratic case, with $\Delta t \propto E^2$, cannot be considered using the same methodology (see, e.g., \cite{AmelinoIJMPD2002_DSR}).}
$\Delta E$ and the source's redshift $z$ was proposed by Jacob \&\ Piran (2008; hereafter \jp) \cite{JacobPiranJCAP2008}. It is based on the ansatz that the local QG physics is the same at each cosmological epoch, while the cosmological evolution is just a background spacetime expansion. It neglects possible local effects due to spacetime curvature on the QG TOF delay.
This formula is used in most of 
astrophysical studies of QG TOF phenomenology. The possibility of an interplay of QG effects and the spacetime expansion, i.e., manifestations of local spacetime curvature, has been discussed in the literature (see \cite{Rosati+PRD2015,AmelinoRosatiBedicPLB2021,Addazi+PrPNP2022,Amelino+JCAP2024} and refs. therein). The effects from such interplay are called curvature-induced since they emerge only in spacetimes with non-zero global curvature. Microscopically, it may be considered as a cosmological evolution of properties of the gravitational field vacuum. Contribution of these effects to $\Delta t(\Delta E, z)$ may be comparable \cite{AmelinoRosatiBedicPLB2021, Amelino+JCAP2024} to that of flat spacetime QG effects, considered within the \jp\ approach.

Including the curvature-induced effects in the $\Delta t (\Delta E, z)$ relation is ambiguous. There are two approaches to QG phenomenology. The first is the broken-symmetry scenario, where the frame invariance symmetry is broken, and the QG vacuum serves as a preferred frame. Within this scenario, curvature-induced effects may lead to almost arbitrary redshift dependence of $\Delta t$ \cite{Rosati+PRD2015,PfeiferPhLB2018}. The second approach is the deformed-symmetry scenario, which keeps the absence of a preferred frame but modifies the transformation and conservation laws to make MDR frame-invariant.
\footnote{In the flat spacetime, the former scenario is often called the Lorentz invariance violation (LIV; the standard model extension \cite{ColladayKosteleckyPRD1998,KosteleckyMewesPRD2009} is the most generic framework for it), while the latter is known as the doubly special relativity (DSR) \cite{AmelinoIJMPD2002_DSR,Amelino+IJMPA2005,Addazi+PrPNP2022}. The applicability of this terminology to expanding spacetimes with curvature-induced QG effects is questionable. Hence, we do not use it here.}
In this case, curvature-induced effects on $\Delta t$ are strongly limited and can be described by a finite number of phenomenological parameters \cite{Rosati+PRD2015,Amelino+JCAP2024}. 

In this work, we compare this finite-parametric form to existing GRB observations and constrain these parameters. In section~\ref{sec:grb_data}, we list the most stringent limits on $\Eqg$, obtained from individual GRBs via the \jp\ formula for $\Delta t(\Delta E, z)$. In section~\ref{sec:curv-ind_formulae}, we describe the formalism for the QG TOF delay developed by Amelino-Camelia et al. (2024; hereafter \ac) \cite{Amelino+JCAP2024} that accounts for curvature-induced QG effects. We show that the limits on $\Eqg$ obtained from the standard analysis can be transformed to the limits on the parameters of the \ac\ formula without reprocessing the observational data. In section~\ref{sec:bayes}, we set rigorous constraints on these parameters using the Bayesian methodology. Finally, we explore, in section~\ref{sec:future}, which future GRB observations may improve the present results. In the concluding section~\ref{sec:concl}, we briefly summarize our results and describe potential caveats of the analysis performed.

\section{Standard gamma-ray burst tests for QG phenomenology}
\label{sec:grb_data}
Assuming the lowest order QG effects to be linear with respect to $E/\Eqg$, for the two photons with energies $E$ and $E+\Delta E$ emitted simultaneously at redshift $z$, the \jp\ approach reads 
\begin{equation}
    \label{eq:jpTOF}
    \Delta t(\Delta E, z) = \sigma\frac{\Delta E}{\Eqg^{(\sigma)}} \int_0^z \frac{(1+z') \diff z'}{H(z')}.
\end{equation}
Here, $\sigma=+1$ for the subluminal MDR (photons with higher energy come later) and $\sigma=-1$ for the superluminal MDR (high-energy photons come earlier). The Hubble constant $H(z)$ is taken from the standard cosmological model \cite{PLANCK_final}:
\begin{equation}
    \label{eq:cosmology}
    H(z) = H_0 \sqrt{\Omega_\text{m}(1+z)^3 +\Omega_\Lambda} = H_0 h(z),
    \quad
    H_0 = 67.4 \frac{\text{km}/\text{s}}{\text{Mpc}},
    \quad
    \Omega_\text{m} = 0.685 = 1 - \Omega_\Lambda.
\end{equation}

A typical TOF delay observational test for the QG phenomenology is estimating of $\Delta t$ and $\Delta E$ for some source with a known $z$, and subsequent constraining of $\Eqg$ from eq.~\eqref{eq:jpTOF}. In this work, we consider only GRBs, as these give the most stringent limits on TOF for linear MDR, which we discuss here.
\footnote{TOF delay tests for other astrophysical sources are listed in, e.g., \texttt{https://qg-mm.unizar.es/wiki/}.}
GRBs have two components: the prompt emission and the afterglow. The prompt emission is predominantly in the low-gamma range, up to tens of GeV. It is highly variable and (in some cases) may consist of short sub-second peaks. The afterglow ranges from radio to {very high energy TeV emission}. It is more steady, with typical variability timescales from seconds to days \cite{PiranRvMPh2005}. The prompt emission provides smaller $\Delta t$ but also smaller $\Delta E$ values, while from the afterglow one can extract larger $\Delta E$ but at larger $\Delta t$ values. Additionally, due to absorption by the extragalactic background light, the TeV afterglow can be detected only from relatively small $z$ sources (see section~\ref{sec:future}).

\begin{table}
    \centering
    \renewcommand{\arraystretch}{1.4}
    \begin{tabular}{lccccccc}
    \hline\hline
        GRB     & type      & Ref.                                         & $z$     & \multicolumn{2}{c}{$\Eqg^{(\sigma)}$ (95\%)}        & \multicolumn{2}{c}{$\kappa$ (95\%)}           \\
                &           &                                              &         & \parbox{2.2cm}{\centering $\sigma = +1$ \\ subluminal} & \parbox{2.2cm}{\centering $\sigma = -1$ \\ superluminal} &    $\kappa_\text{min}$ & $\kappa_\text{max}$       \\
    \hline
        221009A & Afterglow & \cite{PiranOfengeimPRD2024,LHAASO_221009LIV} & $0.151$ & $8.2$                    & $9.0$                    & $-0.017$ & $0.019$  \\
        190114C & Afterglow & \cite{MAGIC_190114LIV}                       & $0.425$ & $0.47$                   & $0.45$                   & $-1.02$  & $0.98$   \\
        090510  & Prompt    & \cite{Vasileou+PRD2013}                      & $0.903$ & $5.2$                    & $11$                     & $-0.091$ & $0.19$   \\
        090902B & Prompt    & \cite{Vasileou+PRD2013}                      & $1.822$ & $0.12$                   & $0.37$                   & $-5.4$   & $17$     \\
        090926A & Prompt    & \cite{Vasileou+PRD2013}                      & $2.107$ & $1.2$                    & $0.17$                   & $-13$    & $1.9$    \\
        080916C & Prompt    & \cite{Vasileou+PRD2013}                      & $4.35$  & $0.22$                   & $0.20$                   & $-21$    & $19$     \\
    \hline\hline
    \end{tabular}
\caption{Our collection of constraints on QG phenomenology from GRB emission analyses within the \jp\ approach (95\% limits). One-sided confidence limits are listed for the sub- and superluminal QG scales $\Eqg^{(\sigma)}$. 
From \cite{Vasileou+PRD2013} and \cite{LHAASO_221009LIV} 
we take the results
for the maximum likelihood method. 
From \cite{MAGIC_190114LIV} 
{we use the results for} the theoretical lightcurve method. 
{The last two columns show the lower and upper limits at the 95\% intervals for  $\kappa$, see eq.~\eqref{eq:kappa-jp} and section~\ref{sec:curv-ind_formulae:est}.}
Note that some of the references use cosmological parameters that are slightly different from those given in eq.~\eqref{eq:cosmology}. This does not introduce a significant error to our estimates. }
    \label{tab:grb_data}
\end{table}

\begin{figure}
    \centering
    \includegraphics[width=0.7\textwidth]{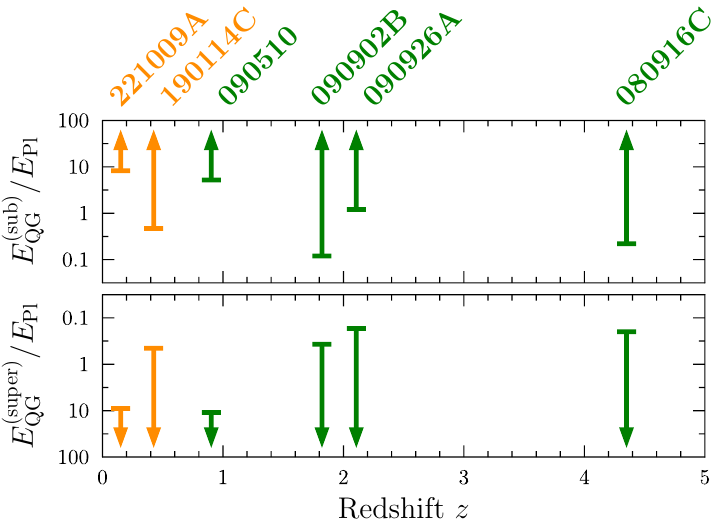}
    \caption{$95\%$-confidence lower limits on the quantum gravity energy scale within \jp\ approach from GRB observations. Orange bars correspond to the results from afterglow TeV emission analyses, and green bars are the constraints of GeV prompt emission. See table~\ref{tab:grb_data} for details.}
    \label{fig:grb_data}
\end{figure}

Today we know several GRBs with TeV afterglow emission detected \cite{MiceliNavaGalaxies2022,LHAASOSci2023afterglow}, but only two of them, 190114C \cite{MAGIC_190114LIV} and 221009A \cite{PiranOfengeimPRD2024,LHAASO_221009LIV}, give significant constraints on $\Eqg$. Observation of the prompt emission must also be of appropriate quality (high temporal resolution and GeV photons detected). Ultimately, the best currently available limits from prompt GRB emission on $\Eqg$ are obtained from four GRBs, 080916C, 090510, 090902B, and 090926A \cite{Vasileou+PRD2013}. We list the properties of these six bursts in table~\ref{tab:grb_data} and depict their QG TOF limits in figure~\ref{fig:grb_data}. 

Before continuing, we point out several ambiguities in the analysis outlined above. First, the observed time lag consists of $\Delta t$ from eq.~\eqref{eq:jpTOF} and some intrinsic lag, which varies for each GRB and has different properties for the prompt emission and the afterglow. The intrinsic lag problem is crucial for constraining $\Eqg$ from an individual source. Nevertheless, simultaneous consideration of several GRBs, especially bringing together the results from prompt emission and afterglow, should resolve this problem. It is inconceivable that intrinsic delays from different bursts from different redshifts will combine to cancel the QG TOF delay. Notice that, despite the diversity of physical processes and cosmological epochs of emission, analyses of the selected six GRBs give comparable (plus-minus an order of magnitude) constraints on the quantum gravity scale $\Eqg \gtrsim (0.1-10) \Epl$.

Second, the observed data can be processed via different methods (pair view, maximum likelihood for specified light curve and spectral model, cross-correlation analysis, etc.) that give different constraints on $\Eqg$. Here, when available,  
we take the results from the maximum likelihood method (see caption to table~\ref{tab:grb_data})

\section{Curvature-induced effects within the symmetry-deformed scenario}
\label{sec:curv-ind_formulae}

\subsection{Formalism}
\label{sec:curv-ind_formulae:formulae}

The method proposed by \ac\ \cite{Amelino+JCAP2024} to study curvature-induced effects on the TOF delay consists of two steps. First, one has to develop the formalism of the deformed-symmetry scenario in the de Sitter universe, the limiting case of expanding spacetime with the Hubble constant being 
exactly constant. The most general deformation of de Sitter algebra of symmetries is five-parametric, but only three independent combinations of them control
$\Delta t$. The second step, denoted de Sitter slicing  \cite{Rosati+PRD2015}, is to 
generalize
the results for the de Sitter universe to a 
universe with a time-dependent Hubble constant, namely the
Friedman-Lema\^{i}tre-Robertson-Walker (FLRW) universe.
\begin{subequations}
The resulting TOF formula
\footnote{
Corresponds to \ac\ notations as $\eta_\lf \leftrightarrow \eta_1$, $\eta_\cin \leftrightarrow \eta_2$, and $\eta_\cia \leftrightarrow \eta_3$.
}
has three terms:
\label{eq:acTOF}
\begin{equation}
    \label{eq:acTOF:Dt}
    \Delta t = \frac{\Delta E}{H_0 \Epl} \left[ 
    \eta_\lf f_\lf(z) + \eta_\cin f_\cin(z) + \eta_\cia f_\cia(z)
    \right].
\end{equation}
The first term that is given by the common \jp\ approach describes the possible QG-induced TOF that can arise even  in the case of the locally-flat spacetime, i.e., with no curvature-induced effects: 
\begin{equation}
    \label{eq:acTOF:f1}
    f_\lf(z) = \int_0^z \frac{(1+z')\diff z'}{h(z')} \ . 
\end{equation}
If one sets $\eta_\cin = \eta_\cia = 0$ and $\eta_\lf = \sigma \Epl/\Eqg^{(\sigma)}$, the TOF formula~\eqref{eq:acTOF:Dt} turns into eq.~\eqref{eq:jpTOF}.
The latter two,
\begin{align}
    f_\cin(z) &= \int_0^z \frac{(1+z')\diff z'}{h(z')}\left[ 1 - \left( 1 - \frac{h(z')}{1+z'}\int_0^{z'} \frac{\diff z''}{h(z'')} \right)^2 \right], 
    \label{eq:acTOF:f2}
    \\
    f_\cia(z) &= \int_0^z \frac{(1+z')\diff z'}{h(z')}\left[ 1 - \left( 1 - \frac{h(z')}{1+z'}\int_0^{z'} \frac{\diff z''}{h(z'')} \right)^4 \right], 
    \label{eq:acTOF:f3}
\end{align}
\end{subequations}
describe the curvature-induced effects.
Hereafter we refer to $\eta_\lf$ as 
the locally-flat parameter, and to $\eta_\cin$ and $\eta_\cia$ as the curvature-induced parameters. 

\subsection{Simple estimates of $\eta_\lf$, $\eta_\cin$, and $\eta_\cia$}
\label{sec:curv-ind_formulae:est}

While the constraints listed in table~\ref{tab:grb_data} are obtained for the \jp\ approach, they can also be used to constrain $\eta_\lf$, $\eta_\cin$, and $\eta_\cia$. Let us introduce the dimensionless quantity
\begin{equation}
    \label{eq:kappa}
    \kappa = H_0 \Delta t \frac{\Epl}{\Delta E} \approx 26.6\frac{\Delta t\,[\text{s}]}{\Delta E\,[\text{GeV}]}.
\end{equation}
If $\Delta t$ and $\Delta E$ are treated as observables for a given GRB, then $\kappa$ quantifies what constraints can be set on the QG parameters from this GRB. Moreover, $\kappa$ is suitable for both \jp\ and \ac\ approaches. Within the former, it is related to the $\Eqg$ scales from table~\ref{tab:grb_data} as
\begin{equation}
    \label{eq:kappa-jp}
    \kappa = \sigma \frac{\Epl}{\Eqg^{(\sigma)}} f_\lf(z).
\end{equation}
Within the latter, the observationally-constrained $\kappa$ is related to the QG parameters as
\begin{equation}
    \label{eq:kappa-ac}
    \kappa = \eta_\lf f_\lf(z) + \eta_\cin f_\cin(z) + \eta_\cia f_\cia(z).
\end{equation}
To constrain the parameters of the \ac\ approach, we 
{perform two steps.} 
First, for each  GRB in our sample, we 
use the limits  available in the literature on $\Eqg$, that were obtained using the common  \jp\ approach, to calculate the limits on $\kappa$. Second, using Eq.~\eqref{eq:kappa-ac} for these $\kappa$ limits we calculate the allowed ranges for $\eta_\lf$, $\eta_\cin$, and $\eta_\cia$.

\begin{figure}
    \centering
    \includegraphics[width=\textwidth]{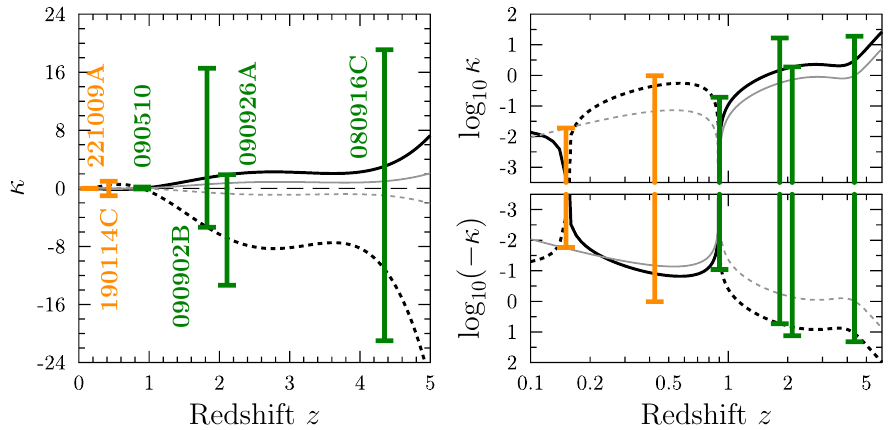}
    \caption{
    Error bars indicating $95\%$ limits on $\kappa$ from the GRB data for the afterglow (orange) and  for the prompt emission (green). The {thick}
    black lines show two limiting examples satisfying all these constraints within the \ac\ approach, solid for $\eta_\lf = 0.52$ and dashed for $\eta_\lf = -1.9$,
    while $\eta_\cin = 14.0\eta_\lf$ and $\eta_\cia = -11.7\eta_\lf$ in both cases. 
    {The thin gray lines show the limiting examples for the case of vanishing locally-flat term, $\eta_\lf = 0$, $\eta_\cin = \pm 2.5$, and $\eta_\cia = \mp 1.8$.}
    The two right panels are the same as the left panel but are log-log-scaled.
    }
    \label{fig:kappas}
\end{figure}

The  $95\%$ limits on $\kappa$ for each of the six GRBs considered here are given in the last column of table~\ref{tab:grb_data} and plotted in figure~\ref{fig:kappas} (left in linear scale, right in log scale). A simple estimate of the limits on the $\eta$ parameters can be obtained by noticing that the $\kappa$ ranges for GRBs 221009A and 090510 are much smaller than those for other bursts. Thus, one can set $\kappa \approx 0$ in eq.~\eqref{eq:kappa-ac} at redshifts of these GRBs, namely $z=0.151$ and $z=0.903$. The resulting two equations make two of the $\eta$ parameters dependent on the third one, e.g.:
\begin{subequations}
\label{eq:eta23of1}
\begin{align}
    \eta_\cin &\approx \frac{f_\cia(0.151) f_\lf(0.903) - f_\cia(0.903) f_\lf(0.151)}{f_\cin(0.151) f_\cia(0.903) - f_\cia(0.151) f_\cin(0.903)} \eta_\lf  \approx 14.0\eta_\lf,
    \label{eq:eta23of1:2}
    \\
    \eta_\cia &\approx  \frac{f_\lf(0.151) f_\cin(0.903) - f_\lf(0.903) f_\cin(0.151)}{f_\cin(0.151) f_\cia(0.903) - f_\cia(0.151) f_\cin(0.903)} \eta_\lf \approx -11.7\eta_\lf.
    \label{eq:eta23of1:3}
\end{align}
\end{subequations}
The range of $\eta_\lf$ values can be obtained by demanding that $\kappa(z)$ given in eq.~\eqref{eq:kappa-ac} satisfies the limits for the remaining four GRBs. At $95\%$ level, this range is 
{$-1.9 \lesssim \eta_\lf \lesssim 0.52$.}
The corresponding ranges of other \ac\ parameters are 
{$-26 \lesssim \eta_\cin \lesssim 7.3$} and 
{$-6.1 \lesssim \eta_\cia \lesssim 22$}.
The two limiting cases are shown in figure~\ref{fig:kappas} by solid and dashed black lines. Surprisingly, the stringent constraints do not arise from GRB~190114C, which is the third in the rank of individual $\kappa$ limits, but from the bursts with higher redshifts: GRB~090926A for the positive side and GRB~090902B for the negative side.
In the following, we obtain limits in a more rigorous way, using a Bayesian analysis.

{This method can be applied also to the case when $\eta_\lf=0$, that is when the locally-flat term vanishes. In this case, $\eta_\cin \propto \eta_\cia$. The constant of proportionality depends on whether we chose GRB 090510 or 221009A as the point in which we set $\kappa \approx 0$. The limits obtained on the curvature-induced terms, that slightly depend on the choice of the anchor point, are $-2.5\lesssim \eta_\cin \lesssim 2.5$ and $-1.8\lesssim \eta_\cia \lesssim 1.8$. The corresponding limiting cases are shown in figure~\ref{fig:kappas} by thin grey lines. Clearly, in view of the strong limits given by the two GRBs 090510 or 221009A the higher redshift points are not significant in this case.}

\section{Bayesian constraining of QG parameters}
\label{sec:bayes}

The Bayes theorem \cite[e.g.,][]{Schweder_Hjort_2016} applied to our problem takes the form
\begin{equation}
    \label{eq:bayes}
    \mathcal{P}(\eta_\lf,\eta_\cin,\eta_\cia | \text{GRB data}) \propto \prod_{i\in\text{GRBs}} \mathcal{L}_i(\text{data}_i | \eta_\lf, \eta_\cin, \eta_\cia) \prod_{k=\lf,\cin,\cia} \pi_k(\eta_k),
\end{equation}
where $\mathcal{L}_i$ is
{the likelihood, i.e.}
the probability 
of the data for the $i$'th GRB yielding a certain $\kappa$ value given the QG parameters $\eta_{\lf,\cin,\cia}$, $\pi_k$ is the prior probability distribution for the $k$'th QG parameter, and $\mathcal{P}$ in the left-hand side is the posterior probability distribution for the QG parameters given the GRB data. The probability distribution for the data (the evidence) is omitted here. We choose the priors $\pi_k(\eta_k)$ being uniform, wide enough so as not to affect the inferences, and symmetric with respect to the sign of each $\eta_k$. For each GRB, we approximate the likelihood as a Gaussian function centered at the middle of the $\kappa$ interval (listed in the last column of table~\ref{tab:grb_data}), with the dispersion adjusted to make this interval the $2\sigma$ credible region for $\kappa$. Thus, for the $i$'th GRB
\begin{equation}
    \label{eq:lihood}
    \mathcal{L}_i = \mathcal{N}\left(\kappa(z_i, \eta_\lf,\eta_\cin,\eta_\cia); \frac{\kappa_{i\,\text{min}}+\kappa_{i\,\text{max}}}{2}, \left[\frac{\kappa_{i\,\text{max}}-\kappa_{i\,\text{min}}}{4}\right]^2\right)
\end{equation}
where $\mathcal{N}(x; \mu, 
\sigma^2)$ is the normal distribution with mean value $\mu$ and dispersion $\sigma$, the function $\kappa(z, \eta_\lf,\eta_\cin,\eta_\cia)$ is defined in eq.~\eqref{eq:kappa-ac}, and $\kappa_{i\,\text{min}}$ and $\kappa_{i\,\text{max}}$ are the lower and upper limits on $\kappa$ for the $i$'th GRB given in table~\ref{tab:grb_data}. 
Note that these likelihood profiles approximate the ones provided in the original papers \cite{Vasileou+PRD2013,MAGIC_190114LIV,LHAASO_221009LIV}.
{Still, this approximation preserves the asymmetry, relative to $\kappa=0$, in the data by taking the mean of the Gaussian distribution to be $\mu_i=({\kappa_{i\,\text{min}}+\kappa_{i\,\text{max}}})/{2}$, that is half way between the upper and lower limits.}
This 
{approximation}
may, of course, affect the constraints we obtain. However, comparing the results 
{(table~\ref{tab:results}) to those presented in section~\ref{sec:curv-ind_formulae:est}, where the upper limits on the QG parameters are dictated solely by the upper error bar of GRB~090926A and the lower ones come from the lower error bar of GRB~090902B,} suggests that those effects are not significant.

\begin{table}
    \centering
    \renewcommand{\arraystretch}{1.4}
    \begin{tabular}{l@{\hspace{0.6cm}}ccc@{\hspace{0.6cm}}rr}
    \hline\hline
                          & \multicolumn{3}{c}{QG parameters ($95\%$)}     & \multicolumn{2}{c}{QG energy scales ($95\%$)} \\
    Case                  & $\eta_\lf$      & $\eta_\cin$  & $\eta_\cia$   & loc.-flat          & curv.-ind.       \\
    \hline
    General               & $-1.8...1.3$    & $-25...18$   & $-15...21$    & $\gtrsim 0.5\Epl$  & $\gtrsim 0.04\Epl$        \\
    Curvature-induced     & $\to 0$         & $-2.6...2.7$ & $-1.9...1.9$  & ---                & $\gtrsim 0.4\Epl$         \\
    Additive energy       & $-0.20...0.18$  & $\to 0$      & $-0.33..0.42$ & $\gtrsim 5\Epl$    & $\gtrsim 2\Epl$           \\
    Locally-flat          & $-0.079...0.10$ & $\to 0$      & $\to 0$       & $\gtrsim 10\Epl$   & ---                       \\
    \hline\hline
    \end{tabular}
    \caption{Constraints on the QG parameters within the \ac\ approach obtained via Bayesian procedure. The second to fourth rows describe special cases for which some of the parameters are set to zero. See text for details.}
    \label{tab:results}
\end{table}

Along with the general case of all three QG parameters being arbitrary, we consider several special cases outlined by \ac. The first one is the purely curvature-induced scenario with no locally-flat QG effects, $\eta_\lf = 0$. Next, we consider the additive energy scenario $\eta_\cin = 0$, when the energy composition law is not affected by the symmetry deformation at the leading order. The third special case is the absence of any curvature-induced effects, $\eta_\cin = \eta_\cia = 0$, which is the \jp\ scenario. In all cases, we sample the posterior distribution for $\eta_\lf$, $\eta_\cin$, and $\eta_\cia$ 
{via Markov chain Monte-Carlo simulations}
using the \texttt{emcee} Python package \cite{emcee2013}, and plot the results using the methods of \texttt{ChainConsumer} \cite{Hinton2016}. The one-dimensional 95\% credibility levels for the QG parameters are summarized in table~\ref{tab:results}. For the sake of connection with other constraints on the QG phenomenology, we define the locally-flat and curvature-induced energy scales
\begin{equation}
    \label{eq:acEscales}
    \Eqg^\text{(lf)} \sim \frac{\Epl}{|\eta_\lf|},
    \quad
    \Eqg^\text{(ci)} \sim \frac{\Epl}{|\eta_{\cin,2}|}.
\end{equation}
Their estimates at 95\% credibility level are listed in the last two columns of table~\ref{tab:results}.

\begin{figure}
    \centering
    \includegraphics[width=\linewidth]{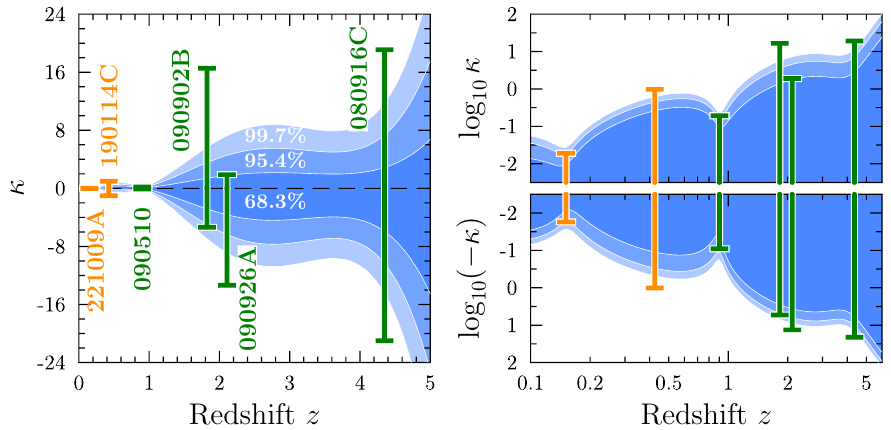}
    \caption{The credibility bands for the $\kappa(z)$ dependence, eq.~\eqref{eq:kappa-ac}, coming from the posterior distributions for the QG parameters in the general case. The vertical error bars for the observed GRBs are the same as in figure~\ref{fig:kappas}. The right plot is the same as the left one but log-log-scaled.}
    \label{fig:kappas+limits}
\end{figure}

In case there are no curvature-induced TOF effects, that is for the purely locally-flat QG case $\eta_\cin = \eta_\cia = 0$, i.e., the \jp\ approach, our fitting procedure yields 95\% limits on the locally-flat parameter $-0.079 < \eta_\lf < 0.10 $.
The corresponding constraint on the energy scale is $\Eqg^\text{(lf)} \gtrsim 10\Epl$. It is  in good agreement with the strongest constraints obtained in previous works \cite{Vasileou+PRD2013,PiranOfengeimPRD2024,LHAASO_221009LIV}. It confirms that the results obtained from different GRBs at different redshifts and from different emission mechanisms (prompt in 090510 and afterglow in 221009A) are consistent with each other.
\footnote{One should not be confused by the fact that the $\kappa$ limits for GRB~221009A are much smaller than those for all other bursts. According to Eq.~\eqref{eq:kappa-jp}, the importance of each GRB for estimating the QG scale is defined by $f_\lf(z)/\kappa$. Table~\ref{tab:grb_data} and figure~\ref{fig:grb_data} show that in the locally-flat case the overall constraint is dominated equally by GRB 090510 and GRB 221009A.}

For the general case, the individual limits on the QG parameters qualitatively match the limits obtained by the simple analysis described in section~\ref{sec:curv-ind_formulae:est}. They are more symmetric since we artificially symmetrize the likelihood functions for individual GRBs, see eq.~\eqref{eq:lihood}. The locally-flat QG energy scale is constrained at the near-Plankian level, $\Eqg^\text{(lf)} \gtrsim 0.5\Epl$, while the curvature-induced energy scale is limited at the sub-Plankian level, $\Eqg^\text{(ci)} \gtrsim 0.04\Epl$. The former constraint is weaker than claimed within the \jp\ approach \cite[e.g.][]{Vasileou+PRD2013,MAGIC_190114LIV,PiranOfengeimPRD2024,LHAASO_221009LIV} due to an obvious reason of larger number of fitting parameters. Using the posterior distribution for the QG parameters and eq.~\eqref{eq:kappa-ac}, we derive the probability distribution of  $\kappa(z)$. For each $z$ it can be treated as a distribution for $\kappa$ values that can be obtained in other observations of a GRB at this redshift. Figure~\ref{fig:kappas+limits} shows 1, 2, and 3$\sigma$ credibility bands of these distributions. 

\begin{figure}
    \centering
    \includegraphics[width=0.49\textwidth]{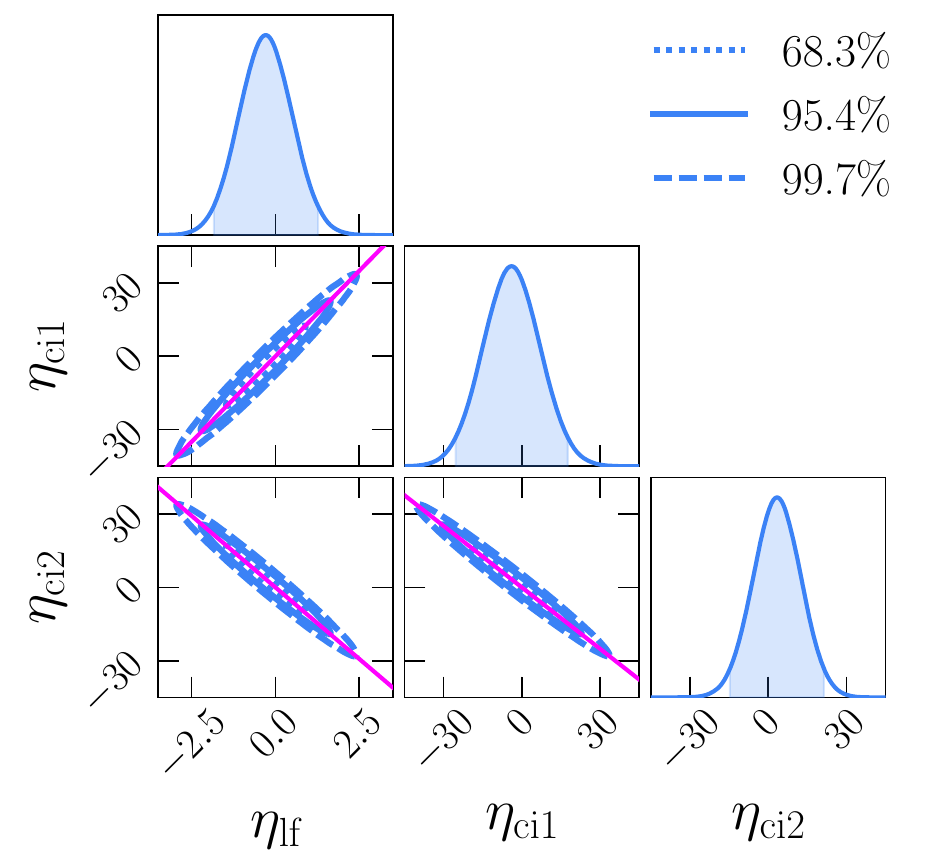}
    \hfill
    \includegraphics[width=0.49\textwidth]{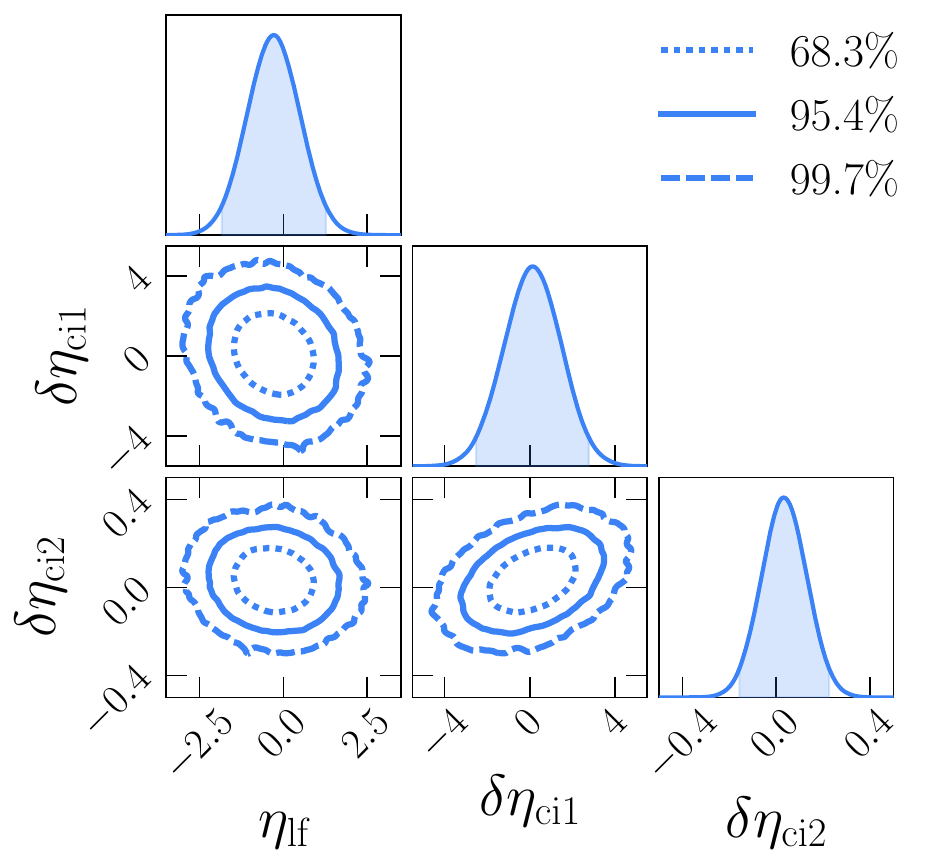}
    \caption{Marginalized one- and two-dimensional posterior distributions for the QG parameters obtained via Bayesian fitting. In the general case, all three parameters are arbitrary. The left panel shows the full range of the parameters, which are strongly correlated. Straight magenta lines show these correlations~\eqref{eq:eta23of1}, that are explained within the naive analysis. On the right panel, the correlations are extracted from the curvature-induced parameters, see eq.~\eqref{eq:uncorr}. Shaded areas under the one-dimensional distributions indicate 95\% credibility intervals.}
    \label{fig:triangle_gen}
\end{figure}

Marginalized one- and two-dimensional posterior distributions of the QG parameters $\eta_\lf$, $\eta_\cin$, and $\eta_\cia$ are shown in the left part of figure~\ref{fig:triangle_gen}. They are strongly correlated. The origin of these correlations is the same as the origin of the relations~\eqref{eq:eta23of1} described in the naive analysis: GRBs~090510 and~221009A set such tight limits on $\kappa$ that the $\kappa(z)$ dependence is ``anchored'' at their redshifts. The magenta lines in the left panel of figure~\ref{fig:triangle_gen} show that eqs.~\eqref{eq:eta23of1} describe the correlations indeed.
Let us emphasize that these correlations are not fundamental and they emerge due to a specific manifold of the QG parameters that satisfy the specific data set. To explore the real room for freedom of the parameters, we use eqs.~\eqref{eq:eta23of1} to extract the correlations from the curvature-induced parameters $\eta_\cin$ and $\eta_\cia$ and define their uncorrelated parts
\begin{equation}
    \label{eq:uncorr}
    \delta\eta_\cin = \eta_\cin - 14.0\eta_\lf, 
    \quad
    \delta\eta_\cia = \eta_\cia + 11.7\eta_\lf + 0.739\delta\eta_\cin.
\end{equation}
Here, we take into account that $\delta\eta_\cin$ is correlated with the part of another curvature-induced parameter remaining after extraction of the correlation with $\eta_\lf$. This part is $\eta_\cia + 11.7\eta_\lf$, and it correlates with $\delta\eta_\cin$ due to the additional ``anchoring'' role of GRB~090510, so $\eta_\cia + 11.7\eta_\lf \approx - \delta\eta_\cin f_\cin(0.903)/f_\cia(0.903) \approx -0.739\delta\eta_\cin$. The triangle plot for $\eta_\lf$, $\delta\eta_\cin$, and $\delta\eta_\cia$ is shown in the right part of figure~\ref{fig:triangle_gen}. While the room for the uncorrelated parts of the curvature-induced parameters is much smaller, the constraints on possible strength of the curvature-induced QG effects are given by the limits on the full parameters $\eta_\cin$ and $\eta_\cia$ and the corresponding energy scale $\Eqg^\text{(ci)}$.

\begin{figure}
    \begin{minipage}{0.49\textwidth}
        \includegraphics[width=\textwidth]{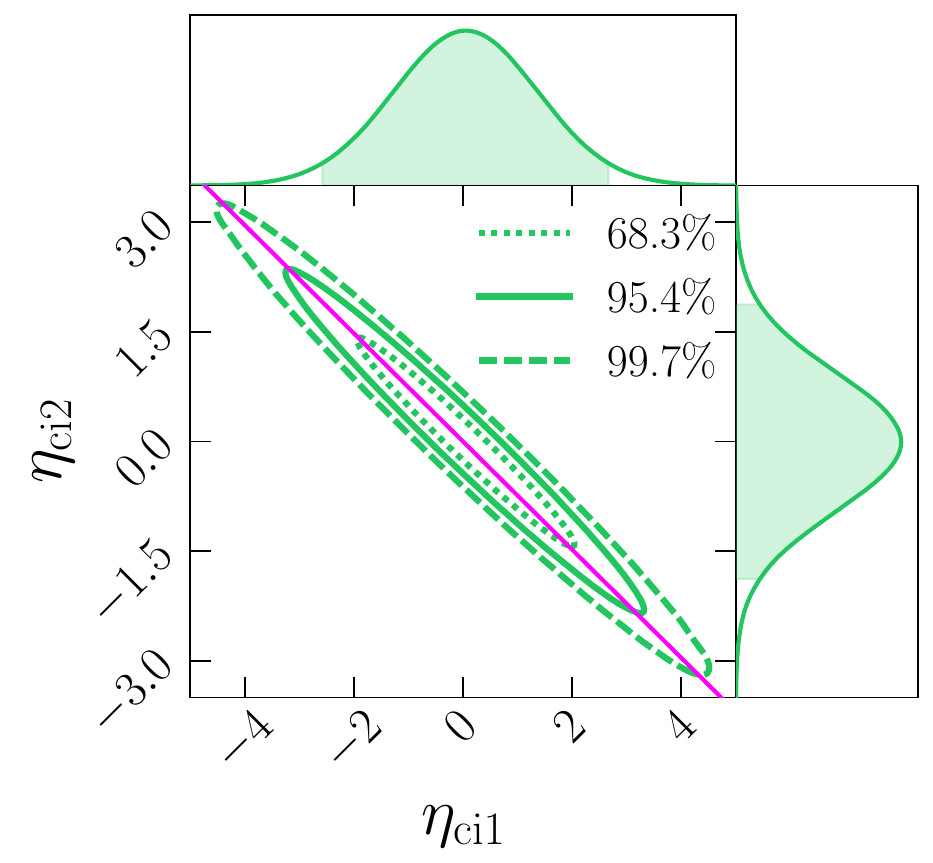}
        \vspace{0.25cm}
    \end{minipage}
    \hfill
    \begin{minipage}{0.49\textwidth}
        \includegraphics[width=\textwidth]{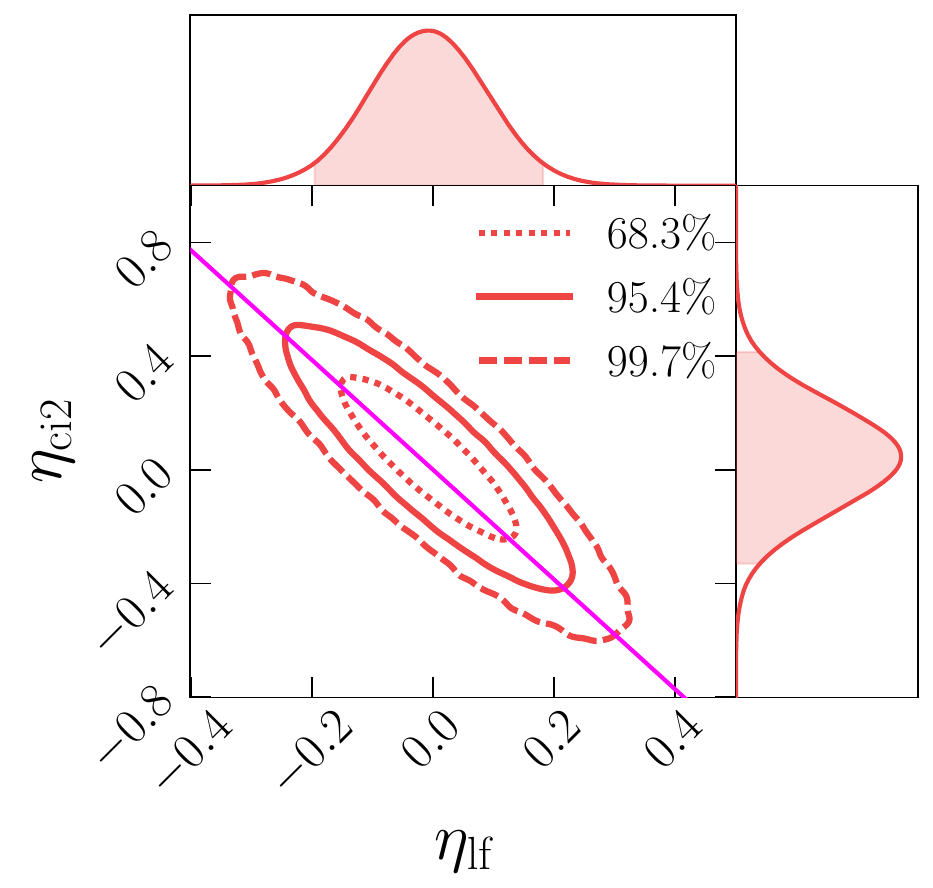}
        \vspace{0cm}
    \end{minipage}
    \caption{Distributions of the QG parameters in cases of the curvature-induced scenario ($\eta_\lf = 0$, left) and the additive energy scenario ($\eta_\cin = 0$, right). Magenta lines show the correlations eqs.~\eqref{eq:eta3of2} and~\eqref{eq:eta3of1}. Shaded areas under the marginalized one-dimensional distributions indicate 95\% credibility intervals.}
    \label{fig:triangle_ci+ae}
\end{figure}

In the specific two-parametric cases, the constraints on the remaining QG parameters are significantly stronger than in the generic case. In the purely curvature-induced scenario, the corresponding energy scale is limited at the near-Plankian level, $\Eqg^\text{(ci)} \gtrsim 0.4\Epl$. This result essentially shrinks the parameter space for phenomenological approaches to QG that have no TOF effect in flat spacetime \cite[e.g.][]{Carmona+U2023}. The additive energy scenario yields the super-Plankian constraints on both locally-flat and curvature-induced scales, $\Eqg^\text{(lf)} \gtrsim 5\Epl$ and $\Eqg^\text{(ci)} \gtrsim 2\Epl$. Figure~\ref{fig:triangle_ci+ae} shows the corresponding two-dimensional distributions. Similarly to the generic case, the QG parameters appear to be strongly correlated. In the curvature-induced scenario, the correlation arises from the anchoring role of GRB~090510 and has the form
\begin{equation}   
    \label{eq:eta3of2}
    \eta_\cia \approx - \frac{f_\cin(0.903)}{f_\cia(0.903)} \eta_\cin \approx -0.739\eta_\cin.
\end{equation}
For the additive energy scenario, GRB~190114C serves as an anchor, and the correlation is
\begin{equation}  \label{eq:eta3of1}
    \eta_\cia \approx - \frac{f_\lf(0.425)}{f_\cia(0.425)} \eta_\lf \approx -1.93\eta_\lf.
\end{equation}
Both relations are shown by magenta lines in figure~\ref{fig:triangle_ci+ae}. As in the generic case, these correlations follow from the specific data set and have no fundamental meaning.

\section{Prospects for future observations}
\label{sec:future}
\begin{figure}
    \centering
    \includegraphics[width=0.8\textwidth]{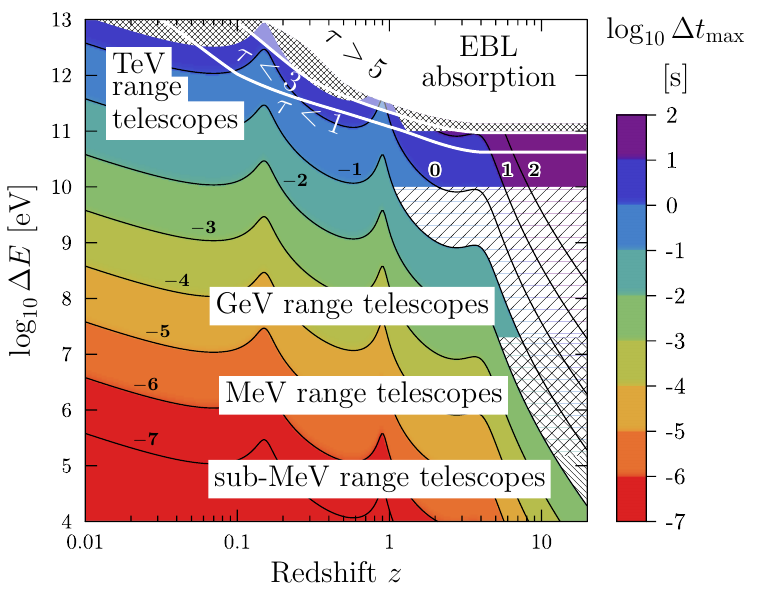}
    \caption{A color map of the maximal value $\Delta t_\text{max}$, eq.~\eqref{eq:Dtmax}, of the TOF delay for photons with energy $\Delta E$ emitted at a redshift $z$ that will improve the constraints on the QG parameters. White lines indicate  $\tau = 1$  and $\tau = 3$ optical depth levels w.r.t. absorption on the extragalactic background light (EBL); the white-shaded region shows the area of $\Delta E$ and $z$ for which  $\tau > 5$. The hatched regions depict energy and time lags that are anticipated to be detectable from GRBs by various telescopes. See text for details.}
    \label{fig:capability}
\end{figure}

The credibility bands on the $\kappa(z)$ dependence (figure~\ref{fig:kappas+limits}) give a hint on which GRBs might be the most promising for further constraining the QG parameters. Suppose that one obtains a new limit on $\Delta t$ for a given $\Delta E$ from a GRB at redshift $z$. 
We can use eq.~\eqref{eq:kappa} to set an upper limit on $|\kappa|$ that characterizes this GRB. This upper limit is informative only if it is stricter than the limits obtained from the current data (i.e. from figure~\ref{fig:kappas+limits}). 
For illustrative purposes, we suppose here that a GRB observation with redshift $z$ can improve these constraints if its $\kappa$ limits are within the 1$\sigma$ bounds shown in  figure~\ref{fig:kappas+limits}. Let us call these bounds $- \kappa_\text{low}^{(1\sigma)}(z)$ and $\kappa_\text{up}^{(1\sigma)}(z)$. Then, one can define $\Delta t$ that will improve the current limits on the QG parameters:
\begin{multline}
    \label{eq:Dtmax}
    \Delta t< \Delta t_\text{max}(\Delta E, z) = \frac{\Delta E}{H_0 \Epl} \max\left\{ \kappa_\text{low}^{(1\sigma)}(z), \kappa_\text{up}^{(1\sigma)}(z) \right\} 
    \\
    \approx 0.0376\,\text{s}\, \frac{\Delta E}{\text{GeV}} \max\left\{ \kappa_\text{low}^{(1\sigma)}(z), \kappa_\text{up}^{(1\sigma)}(z) \right\} \ .
\end{multline}
The function $\Delta t_\text{max}(\Delta E, z)$ is shown in figure~\ref{fig:capability}. Obviously, the required $\Delta t$ constraints are smaller for larger energies and redshifts, and they are larger near redshifts close to those of GRBs~221009A and~090510 (seen as the two peaks of $\Delta t_\text{max} = \text{const}$ profiles). 

Now we can confront $\Delta t_\text{max}(\Delta E, z)$ with capabilities of various 
telescopes that detect GRBs and their afterglows. A telescope is schematically characterized by the $\Delta E$ range and the lower limit of measurable $\Delta t$. 
Both these quantities are determined by a combination of detector characteristics and GRB signatures. The former is determined by the detector channels and physical limitations of the detected emission type, while the latter comes partially from the telescope timing resolution and mainly from the variability of the detected emission.
A necessary condition for a GRB to improve the QG limits is that the minimal $\Delta t$ one can measure with the telescope
is smaller than
$\Delta t_\text{max}(\Delta E, z)$.

Observations by the \textit{Fermi} observatory that had a significant impact on constraining the QG phenomenology \cite{Vasileou+PRD2013} were performed by the \textit{Fermi}-LAT telescope. It can measure $\Delta E$ roughly between $20\,$MeV and $300\,$GeV \cite{FermiLAT2009}, but if we focus on the prompt emission, we should lower the upper limit to about $10\,$GeV as prompt photons above this energy have been rarely detected. The minimum $\Delta t$ that can be detected for GRBs in this energy range can be roughly understood from the limits { obtained } from the analysis of GRB~090510 \cite{Vasileou+PRD2013}. We estimate it as $0.1\,$s.
\footnote{
Note that this is far above the timing resolution of the detector.
} The area satisfying two conditions $20\,\text{MeV} < \Delta E < 10\,\text{GeV}$ and $\Delta t_\text{max}(\Delta E, z) > 0.1\,$s is sparsely single-hatched in figure~\ref{fig:capability}. A \textit{Fermi} detection of a GRB at a redshift $z\gtrsim 1$ in a GeV range can improve the constraints on the QG parameters if the observed $\Delta t$ is small enough. Another option is to detect a GRB in the range of tens of MeV at redshift $z\sim 10$, but it is a much less probable event.

Similarly, we consider the capabilities of the sub-MeV range telescopes such as \textit{Swift}-BAT \cite{SwiftBAT2005} (also Konus-Wind \cite{Konus-Wind}, Insight-HXMT \cite{Insight-HXMT}, etc.). Its energy channels allow it to measure $\Delta E$ up to $150\,$keV, and the time lags it can detect in practice \cite{Ellis+ApP2006} are $\Delta t \gtrsim 0.01\,$s. The corresponding area in the $\Delta E - z$ plane is densely single-hatched in the right bottom corner of figure~\ref{fig:capability}.  A  sub-MeV telescope has to detect a GRB at a redshift greater than $10$ to make a significant contribution to constraining the QG phenomenology. This is unlikely given the rareness of such events and the required limits on $\Delta t$.

The range between sub-MeV and GeV is covered by MeV-range telescopes like \textit{Fermi}-GBM or newly-lanched \textit{SVOM}-GRM. Regardless of their timing resolutions, we estimate the strongest constraint they can set on $\Delta t$ as about $0.01\,$s. This estimate comes from the properties of the prompt emission of GRBs, whose variability is rarely faster than a millisecond in the rest frame and is redshifted in the detector frame. The corresponding area in figure~\ref{fig:capability} is sparsely cross-hatched. Even if these telescopes detect a GRB near the upper verge of the energy band with a good signal-to-noise ratio, it has to have a high redshift ($z > 5$) to constrain the QG effects valuably. 

We base our estimates of the capabilities of TeV range telescopes (e.g. LHAASO \cite{LHAASOSci2023afterglow}, MAGIC \cite{MAGIC_190114LIV}, H.E.S.S. \cite{HESS}) on the LHAASO observations of the GRB~221009A afterglow~\cite{LHAASOSci2023afterglow}. Namely, we suppose that observations by such telescopes can yield $\Delta E > 100\,$GeV and $\Delta t$ larger than a few seconds. This area is densely cross-hatched in figure~\ref{fig:capability}. Here one has to take into account the absorption of $\gamma$-photons by the extragalactic background light (EBL). This strongly limits the distance up to which events can be detected. We depict the levels of optical depth for this process $\tau = 1$, $3$, and $5$ according to the EBL measurements by ref.~\cite{FranceschiniU2021}. Ultimately, TeV range telescopes can improve our constraints on $\eta_\lf$, $\eta_\cin$, and $\eta_\cia$ if they detect either $10\,$TeV photons from an afterglow of a nearby GRB with $z<0.1$, or TeV and sub-TeV photons from a GRB with large redshifts.

\section{Conclusion}
\label{sec:concl}

This work, for the first time, provides a detailed investigation of the constraints on the linear locally-flat and curvature-induced (within the symmetry-deformed scenario \cite{Amelino+JCAP2024}) QG TOF effects. Using the \ac\ formalism, we explored the parameter space for both locally-flat and curvature-induced contributions, setting 95\% credibility bounds on all three terms of the generic TOF delay formula. Using observations of the six most suitable GRBs and Bayesian analysis, we find that curvature-induced TOF effects may only emerge at energy scales exceeding $0.04\Epl$. Because of the additional degrees of freedom introduced, when curvature-induced effects exist, the limit on the locally-flat QG scale is $0.5\Epl$, as compared to $\sim 10\Epl$ when these effects don't exist.

If one considers the special case of purely curvature-induced TOF delay with the absence of locally-flat effects, then the limit on the former energy scale increases by a factor of 10. This significantly limits the room for the QG phenomenology theories, which have no TOF delay in locally-flat spacetime \cite{Carmona+U2023}. A byproduct of our analysis is the confirmation of the $\sim 10\Epl$ constraint on the locally-flat QG scale using a combined test of multiple GRBs and different emission processes. This strongly enhances the reliability of this limit (see, e.g., \cite{Ellis+PRD2019}). 

Our findings suggest that future observations can improve these constraints. 
{Particularly, increasing the number of GRBs suitable for the analysis will make the QG TOF limits more robust with respect to the intrinsic effects problem.}
We identify specific redshift-energy combinations that would maximize the sensitivity to QG parameters. For instance, observations of GeV GRBs, like GRB~090510, at moderate redshifts, or nearby TeV GRBs with minimal extragalactic background light absorption could provide valuable insights. However, it is unlikely that the limit of $10 \Epl$ for the locally-flat scale can be improved significantly, as those limits are stretching the capability of both GeV and TeV range telescopes.

\acknowledgments

We thank P.S.~Shternin  and S.A.~Balashev for helpful advice concerning Bayesian analysis. This work was supported by Advanced ERC grant MultiJets.

\bibliographystyle{JHEP}
\providecommand{\href}[2]{#2}\begingroup\raggedright\endgroup

\end{document}